\documentclass[draft]{agujournal2019}
\usepackage{url} 
\usepackage{amssymb}
\usepackage{lineno}
\usepackage[finalnew]{trackchanges}
\addeditor{P.W.}
\usepackage{soul}

\draftfalse

\journalname{JGR: Planets}

\begin{document}

\title{The Effects of a Stably Stratified Region with radially varying Electrical Conductivity on the Formation of Zonal Winds on Gas Planets}

\authors{P. Wulff\affil{1}\affil{2}}
\authors{U.~R. Christensen\affil{1}}
\authors{W. Dietrich\affil{1}}
\authors{J. Wicht\affil{1}}

\affiliation{1}{Max Planck Institute for Solar System Research, G\"ottingen, Germany}
\affiliation{2}{Georg-August University, G\"ottingen, Germany}

\correspondingauthor{Paula Wulff}{wulff@mps.mpg.de}

\begin{keypoints}
\item Our numerical models examine the conditions needed to form zonal winds in gas planets, complying with observed gravity and magnetic data
\item A stable layer and magnetic forces are key for strong surface winds at high latitudes that are damped at the inferred depths
\item The decay profile of the winds in the stable layer is controlled by the product of conductivity and squared magnetic field strength
\end{keypoints}

\begin{abstract}
The outer areas of Jupiter and Saturn have multiple zonal winds, reaching the high latitudes, that penetrate deep into the planets' interiors, as suggested by gravity measurements. These characteristics are replicable in numerical simulations by including both a shallow stably stratified layer, below a convecting envelope, and increasing electrical conductivity. A dipolar magnetic field, assumed to be generated by a dynamo below our model, is imposed. We find that the winds' depth into the stratified layer depends on the local product of the squared magnetic field strength and electrical conductivity. The key for the drop-off of the zonal winds is a meridional circulation which perturbs the density structure in the stable layer. In the stable region its dynamics is governed by a balance between Coriolis and electromagnetic forces. Our models suggest that a stable layer extending into weakly conducting regions could account for the observed deep zonal wind structures.
\end{abstract}

\section*{Plain Language Summary}
Jupiter and Saturn's atmospheres display persistent east-west zonal jets, similar to Earth. These jets, extending 2,500-3,000~km and 8,000-9,000~km into Jupiter and Saturn's interiors respectively, have been challenging to simulate. Current numerical models struggle to replicate multiple jets, spanning all latitudes and their decay at the depths inferred from gravity measurements. This study explores the hypothesis that a stably stratified layer, located at the transition to a semi-conducting region, allows the generation of mid-latitude zonal winds and their damping at depth. Using 3D numerical simulations, we model the outer 30\% of the planets where the upper part convects and the lower part is stably stratified. We impose a dipolar magnetic field at the lower boundary and electrical conductivity increases with depth. We observe that the decay in jet amplitude in the stable region depends on the local strength of the magnetic forces. Deep within the stable region, these Lorentz forces are balanced by meridional flow, which leads to temperature perturbations and efficient zonal wind quenching.

\section{Introduction}\label{sec:Introduction}

Zonal winds are alternately westwards/eastwards flows and feature across all four outer planets in our solar system. Those observed on the gas giants, Jupiter and Saturn, share some key characteristics. The dominating equatorial prograde flow on Jupiter (Saturn) spans roughly $30^\circ$ ($60^\circ$) with an amplitude of around $100$~m/s ($400$~m/s). This is flanked by a pair of slightly weaker retrograde jets and multiple jets reaching the high-latitude regions \cite{Tollefson_2017, GarciaMelendo_2011}. While these winds are weaker, they are still significantly stronger in amplitude than non-zonal flows. Jupiter's northern hemisphere also features an unusual prograde jet, as strong as the equatorial jet, at around $21^\circ$ latitude, introducing a strong equatorial antisymmetry into the dynamics.\\
Surface measurements, first from Voyager 1 and 2 \cite{Ingersoll_1981} then from Cassini \cite{Salyk_2006}, have shown a strong correlation between the eddy momentum flux (or Reynolds stresses) and the zonal wind speed as a function of latitude. This confirms current theories that Reynolds stresses, which are statistical correlations of the components of the flow at small and intermediate scales, drive the zonal winds.\\
The extent of the winds into the jovian interior has recently been constrained using the gravity moment measurements from Juno, yielding a depth between $2,500-3,000$km; around $96\%$ of the planet's radius \cite{Kaspi_2018, Dietrich_2021, Galanti_2021}. The same investigation has also been carried out for Saturn, using the Cassini measurements, suggesting the winds extend to $8,000-9,000$km depth, around $85\%$ of the planetary radius \cite{Galanti_2019}. \\
This is consistent with simulation-based studies  where it has been found that the location of the flanking retrograde jets is usually coincident with the `tangent cylinder', here loosely defined as the cylinder aligned with the axis of rotation with a radius corresponding to the depth at which jet quenching takes place. This has been found in numerical models studying both magnetic effects as a potential braking mechanism for the winds, with increasing electrical conductivity at depth (eg. \citeA{Heimpel_2011, Duarte_2013}) or transition into a stably stratified region \cite{Wulff_2022}. Therefore, based on these basic geometric observations of the dynamics we would expect the winds to penetrate deeper on Saturn, with its much wider equatorial jet.\\
A strong prograde jet flanked by two retrograde jets in the equatorial region, outside the tangent cylinder, were already reproduced in hydrodynamic simulations \cite{Christensen_2002, Heimpel_2005, Gastine_2014}. The latter two studies, with larger aspect ratios even exhibited some zonal flow inside the tangent cylinder. However, these models had stress-free inner boundary conditions and therefore failed to provide any insights into the winds' damping mechanism in the interior.\\
In both planets the increasing electrical conductivity at depth (e.g. \citeA{French_2012}), plays a crucial role in the zonal winds' downward propagation from the surface. It has been speculated that deeply penetrating zonal winds may cause the observed secular variation \cite{Moore_2019}. However, \citeA{Bloxham_2022} argue that a slight correction of Jupiter's rotation rate provides a better explanation, in combination with deeper flows in the dynamo region. Furthermore, considering reasonable limits for the total ohmic dissipation suggests that the winds may not penetrate into the highly conducting region of Jupiter \cite{Liu_2008, Wicht_2019, Cao_2017}. It was originally surmised that Lorentz forces, acting where the deep zonal flows reach the conducting region, were responsible for the braking of the winds. However, simulation-based studies such as \citeA{Dietrich_2018} found that these Maxwell stresses at depth eradicate all large scale zonal flow above the conducting region, leading to zonal wind profiles with the strong flows confined to near the equator. \\
\citeA{Christensen_2020} suggested that a combination of a stably stratified layer (SSL) and the magnetic effects at depth are responsible for the braking of the zonal flows on Jupiter. They suggest that the winds decrease in the stable layer in accord with a thermal wind balance. The required density perturbation is caused by a meridional circulation which is affected by electromagnetic forces. \citeA{Duer_2021} present observational evidence for the existence of meridional flow associated with the winds. \citeA{Gastine_2021} conducted a global dynamo simulation with a strong radial variation of conductivity, which was successful in producing winds formed and being maintained above the highly electrically conducting region. Recently, \citeA{Moore_2022} also showed that dynamo simulations of Jupiter including a SSL at $90-95\%$ radius produced dynamos with a dominant axial dipole component and a similar degree of complexity as the measured Jovian magnetic field.\\
In the context of Saturn a stable layer, shallower than the region of metallic conductivity, could help to explain both the formation of its high-latitude zonal winds and how they are quenched at depth, and its magnetic field. This is remarkably axisymmetric \cite{Dougherty_2018} and a stable layer at the top of its semi-conducting region would provide a skin-effect, reducing the smaller-scale field components (suggested by \citeA{Stevenson_1979} and studied by \citeA{Christensen_2008, Stanley_2008, Stanley_2010}). Furthermore, the difference in amplitude of its axial dipole field compared to the higher degree $m=0$ components \cite{Cao_2020} indicates that there may be both a deeper dynamo region generating the strong dipole field, located between a dilute core and the helium rain layer, as well as a shallower layer adding the weaker latitudinally banded perturbations, operating between the helium rain region and a shallower, thin, stable layer. \\
However, for both planets the main uncertainty in the hypothesis is the origin, location, depth and strength of such a relatively shallow stable layer. A helium rain layer \cite{Stevenson_1977}, providing a potential source of compositional stratification, is predicted to lie deeper than the extent of the zonal winds. In Jupiter, although there are some uncertainties concerning the H/He phase diagram, this would be below $86\%$ radius based on \textit{ab initio} EoS calculations of high-pressure experiments \cite{Hubbard_2016, Lorenzen_2011, Brygoo_2021}. In Saturn helium immiscibility may occur at around $65\%$ radius, e.g. \citeA{Morales_2013}. In both planets, however, there is not only a large uncertainty with regards to the depth of a helium rain layer but also no good estimate for its vertical extent. For the case of Jupiter the shallower regions, above where a helium rain layer is thought to reside, are potentially also more complex, based on the accurate gravity measurements from Juno, which suggests the existence of a shallow stably stratified region \cite{Debras_2019, Nettelmann_2021, Debras_2021}, providing a potential link with the stable region associated with a quenching of the zonal winds. \\
In \citeA{Wulff_2022} we used purely hydrodynamic convection models to investigate the relationship between the degree of stratification of such a layer and the penetration of the winds, formed in the overlying convecting envelope, into the stable region below. We found that when the degree of stratification is strong, zonal flows form all the way to the higher latitudes, as is observed on both gas giants, even when imposing a no-slip boundary condition at the bottom of the stable layer. Furthermore, when encountering the SSL, the winds are quenched and geostrophy (i.e. their invariance with respect to the axis of rotation) is broken. However, the decay of the jet amplitude in this hydrodynamic study was still too gradual with depth to fit secular variation data. Furthermore, we expect that at sufficient depth the electrical conductivity will be large enough for magnetic effects to play a role. Therefore, it is crucial to investigate how this will influence both the damping of the jets in the SSL as well as their strength and latitudinal distribution in the overlying convective region. In our study we also test the concept of \citeA{Christensen_2020}. In their simplified models the zonal flow was driven by an imposed ad-hoc force. In our models the zonal winds are driven self-consistently by the convective eddies, which implies that a potential feedback of the winds on the eddy dynamics is also accounted for.\\
\section{Methods}\label{sec:Theory}

We simulate thermal convection in a spherical shell rotating with angular velocity $\Omega\cdot\mathbf{\hat{e}}_z$. The ratio of inner boundary radius, $r_i$, to outer radius, $r_o$, is 0.7. Only the upper part of the shell above $0.83r_o$ is convectively unstable, whereas the lower part is stably stratified (described in detail in Section~\ref{sec:SSL}). We assume an exponentially varying electrical conductivity rising from a negligible value at $r_o$ to a moderate value at $r_i$ (see Section~\ref{sec:MagDiffusivity}). We impose an axisymmetric dipolar magnetic field aligned with the rotation axis through a boundary condition at $r_i$, which represents a field generated by a dynamo operating below $r_i$. For our systematic study we use the Boussinesq approximation (i.e. incompressible flow), although we also perform additional simulations with the anelastic approximation (where a radially varying background density is prescribed). The Boussinesq simulations are cheaper computationally and allow a wider parameter study. In this study, we keep all hydrodynamic parameters as well as the degree of stability in the SSL at fixed values, but we vary the magnetic field strength and the profile of the electrical conductivity. The anelastic simulations are carried out for a subset of these parameters in order to confirm that the trends observed also hold in the compressible models.\\

\subsection{MHD Equations}\label{sec:MHDequations}

As our primary analysis focuses on simulations that use the Boussinesq approximation, we give the governing magneto-hydrodynamic (MHD) equations here in their incompressible form (see \citeA{Wulff_2022} for the hydrodynamic equations under the anelastic approximation). The key features we incorporate are the radially varying magnetic diffusivity $\lambda(r)$ and $dT_c/dr$, the imposed stratification profile, where $T_c$ is the background temperature. As we use a constant gravity, $g$, the equations then simplify to:
\begin{eqnarray}
    \frac{\partial \mathbf{u}}{\partial t} + (\mathbf{u}\cdot\nabla)\mathbf{u} + \frac{2}{E}\mathbf{\hat{e}}_z\times\mathbf{u} & = & -\nabla p + \frac{Ra}{Pr} \vartheta \mathbf{\hat{e}}_r + \frac{1}{E Pm} (\nabla\times\mathbf{B})\times\mathbf{B} + \nabla^2 \mathbf{u}, \\
    \frac{\partial\mathbf{B}}{\partial t} & = & \nabla\times(\mathbf{u}\times\mathbf{B}) - \frac{1}{Pm}\nabla\times(\lambda(r)\nabla\times\mathbf{B}),\label{eq:Bfield}\\
    \frac{\partial \vartheta}{\partial t} + (\mathbf{u}\cdot\nabla)\vartheta + u_r\frac{dT_c}{dr} & = & \frac{1}{Pr}\nabla^2 \vartheta, \\
    \nabla\cdot\mathbf{u} & = & 0,\\
    \nabla\cdot\mathbf{B} & = & 0,
\end{eqnarray}
where $\mathbf{u}$ is the velocity field, $\mathbf{B}$ is the magnetic field, and $p$ is pressure. Temperature fluctuations $\vartheta$ are defined with respect to the hydrostatic reference state. We adopt a dimensionless formulation where the reference length scale is the shell thickness $d=r_o-r_i$, where $i$ denotes the inner boundary values and $o$ denotes outer boundary. Time is given in units of the viscous diffusion time $\tau_\nu=d^2/\nu$, where $\nu$ is the fluid viscosity. The temperature scale is normalised by the value of the gradient of the background temperature at the outer boundary $|dT_c/dr|_o$, multiplied by $d$ (see \citeA{Gastine_2020} for a Boussinesq study involving a stable layer implemented in a similar way). The non-dimensionalised velocity is equivalent to a Reynolds number $Re=ud/\nu$. The magnetic field is given in units of $\sqrt{\rho_o\mu\lambda_i\Omega}$, where $\mu$ is the magnetic permeability and $\lambda$ is the magnetic diffusivity which we prescribe as an analytical radial profile.\\
The dimensionless control parameters that appear in the equations above are the Ekman number ($E$), Rayleigh number ($Ra$), Prandtl number ($Pr$) and magnetic Prandtl number ($Pm$). They are defined as
\begin{equation}
    E=\frac{\nu}{\Omega d^2}, \quad Ra=\frac{\alpha g d^4}{\kappa\nu}\left|\frac{dT_c}{dr}\right|_o, \quad Pr = \frac{\nu}{\kappa}, \quad Pm = \frac{\nu}{\lambda},
\end{equation}
where $\kappa$ is the thermal diffusivity and $\alpha$ is the thermal expansivity. The magnetic Prandtl number $Pm$, based on a reference value of the magnetic diffusivity, is kept at 0.5. However, the magnetic diffusivity at the lower boundary $\lambda_i$ is varied.\\
The mechanical boundary conditions are stress-free at both $r_i$ and $r_o$ and we apply fixed entropy at the outer boundary and fixed entropy flux (downward in our models) at the inner boundary.

\subsection{Hydrodynamic Control Parameters}\label{sec:Variables}

We perform our simulations at a (nominal) Rayleigh number $Ra=6\times10^8$, Ekman number $E=10^{-5}$ and Prandtl number $Pr=0.5$. This yields a convective Rossby number of:
\begin{equation}
    Ro_c = E \sqrt{Ra/Pr} = 0.346.
\end{equation}
Ensuring that this is below 1 helps us to aim for a regime that leads to the formation of a prograde equatorial jet, based on \citeA{Gastine_2013}.\\
Some additional simulations are carried out under the anelastic approximation (see \citeA{Wulff_2022} for the governing equations), with polytropic index 2 and dissipation number 1.5, yielding a mild density stratification of $\rho(r_i)/\rho(r_o)=6.25$. This choice is motivated by aiming for a similar density contrast across the stable layer as one may expect for Jupiter. Our SSL lies at $\sim0.57$ dimensionless depth. Taking $r_c\approx0.97r_J$ for Jupiter, then $r_i$ would lie at $0.948r_J$, keeping the same proportions between convecting and stable layer as in our model. Using a Jovian density profile from \citeA{Guillot_2004} this gives $\rho(r_i)/\rho(r_c)\approx1.83$ which is matched with our choice of the dissipation number. From \citeA{Jones_2009} and \citeA{Gastine_2012}, for example, we know that the critical Rayleigh number increases with increasing density stratification. From the latter study we estimate that the increase is roughly two-fold, compared to our Boussinesq models. Therefore, to compare simulations with a similar degree of supercriticality, we double $Ra$ for the anelastic simulations.\\
The values given above are based on the full shell width $d$. Table \ref{tab:Rescaled} also lists both non-dimensional numbers, re-scaled to the thickness of the convective region (the outer $\sim57\%$). We also give the Rayleigh number based on the temperature (entropy for the anelastic cases) difference across the convective region alone, calculated from the horizontally averaged temperature (entropy) drop across the convecting region:
\begin{equation}\label{eq:RaDelta}
    \textrm{Boussinesq: } Ra_\Delta = \frac{\alpha g (r_o-r_c)^3 \Delta T}{\kappa \nu}, \quad \textrm{Anelastic: } Ra_\Delta = \frac{\alpha g T_o (r_o-r_c)^3 \Delta s}{c_p \kappa \nu},
\end{equation}
where $r_c$ is the bottom of the convective region.
\begin{table}
\caption{Nominal Ekman and Rayleigh numbers are based on the full shell thickness and the surface entropy flux. Their re-scaled values, $E_c$ and $Ra_c$, are based on the thickness of the convective region $d_c=r_o-r_c\approx0.57$. The re-scaled Ekman number for the stable region $E_s$ is also given, based on $d_s=r_c-r_i\approx0.43$. $Ra_\Delta$ is the Rayleigh number defined by Eq.~\ref{eq:RaDelta}. $A_s$ is the value of $dT_c/dr$ ($d\tilde{S}/dr$ for anelastic cases) in the stable region.\label{tab:Rescaled}}
\centering
\begin{tabular}{ c c c c c c c c}
    \hline
    Sim. & $E$ & $E_s$ & $E_c$ & $Ra$ & $Ra_c$ & $Ra_\Delta$ & $A_s$ \\
    \hline
    H & $10^{-5}$ & $5.24\times10^{-5}$ & $3.15\times10^{-5}$ & $6\times10^8$ & $6.04\times10^7$ & $3.6\times10^7$ & 200 \\
    B & $10^{-5}$ & $5.24\times10^{-5}$ & $3.15\times10^{-5}$ & $6\times10^8$ & $6.04\times10^7$ & $2.6\times10^7$ & 200 \\
    A & $10^{-5}$ & $5.24\times10^{-5}$ & $3.15\times10^{-5}$ & $1.2\times10^9$ & $1.21\times10^8$ & $6.2\times10^7$ & 100
\end{tabular}
\end{table}

\subsection{Stably Stratified Layer}\label{sec:SSL}
\begin{figure}
\noindent\includegraphics[width=15cm]{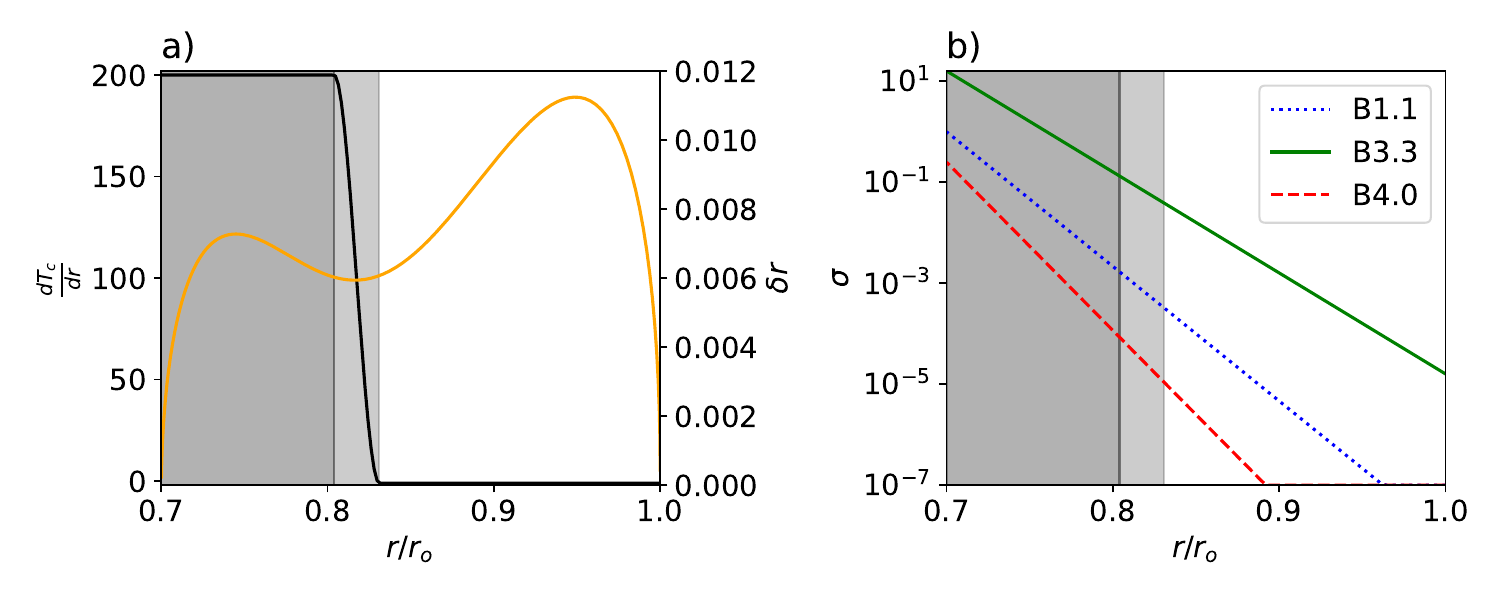}
\caption{a) $dT_c/dr$ profile (black) described in Section ~\ref{sec:SSL}. The grey shaded region indicates $r_s<r<r_c$ while the dark grey region is fully stratified. The radial grid-point separation is shown in orange (right y-axis). b) electrical conductivity $\sigma=\lambda^{-1}$ for reference case B1.1 (blue) and extreme cases B3.3 (green) and B4.0 (red).}
\label{fig:RadDist}
\end{figure}

The region $r>r_c$ is fully convective, whereas at $r<r_s$ the full degree of stability has been reached, with a transition region at $r_s<r<r_c$. This is implemented by prescribing an analytic background entropy gradient profile defined, using auxiliary variable $\chi=(r-r_c)/(r_c-r_s)$, by:\\
\begin{math}
    \frac{dT_c}{dr} =\left\{
        \begin{array}{ll}
        A_s, & \mbox{if $r\le r_s$},\\
        (A_s+1)\cdot\chi^2 \cdot(2\chi + 3) - 1, & \mbox{if $r_s<r<r_c$},\\
        -1, & \mbox{if $r\ge r_c$}.
        \end{array}
    \right.
\end{math}\\
This is plotted in Figure~\ref{fig:RadDist}a. Having investigated the effects of varying the degree of stability in \citeA{Wulff_2022}, we keep the properties of the SSL the same for all cases in this study: $r_c=2.77=0.831r_o$, $r_s=2.68=0.804r_o$ and $A_s=200$ ($A_s=100$ for the anelastic cases). Neutral stability is reached around $0.830r_o$. The ratio of the Brunt-V\"{a}is\"{a}l\"{a} frequency, $N$, to the rotation rate, is:
\begin{equation}
    N/\Omega = \sqrt{\frac{RaE^2}{Pr}A_s},
\end{equation}
which is equal to 4.9, at $ r\le r_s$. This quantifies the effect of the restoring buoyancy force relative to the rotational forces and corresponds to a degree of stratification around the middle of the range studied in \citeA{Wulff_2022}. This parameter is kept the same for the anelastic cases. Comparing this with the range $N/\Omega=2-7$ from \citeA{Mankovich_2021} for a diffuse Saturnian core, may suggest that our value is plausible in terms of its order of magnitude.

\subsection{Magnetic Parameters}\label{sec:MagDiffusivity}

We vary the magnetic diffusivity $\lambda$, or the electrical conductivity $\sigma=1/\lambda$, in this study but keep all other diffusivities ($\nu$ and $\kappa$) constant. We prescribe the magnetic diffusivity to be:\\
\begin{equation}\label{eq:MagDiff}
    \lambda = \lambda_i \exp\left(\frac{1}{d_\lambda} (r - r_i)\right).
\end{equation}
For the profiles where $\lambda$ would exceed $10^7$ we cap it at this value to avoid numerical problems. The electrical conductivity scale height is $d_\lambda=[(1/\lambda)\cdot d\lambda/dr]^{-1}$. This simple exponential profile gives the convenience of having a constant scale height throughout the shell.\\
In our reference model $\lambda_i=1$ and $d_\lambda=1/\ln(10^8)\approx0.054$. To investigate and distinguish the effects of a different local value of electrical conductivity and a different scale-height, we vary both $d_\lambda$ and $\lambda_i$ in a systematic parameter study  (see Table~\ref{tab:Cases}). The electrical conductivity profiles of the extremes of the study, B3.3 and B4.0, are shown in Figure~\ref{fig:RadDist}b).\\
An axial dipole field (poloidal $\ell=1$, $m=0$ component) with amplitude $B_{dip}$ at the poles is imposed as a boundary condition at $r_i$ (negative at the North pole). The other poloidal components and the toroidal field are matched to a field in the inner core, obtained by solving the induction equation in the inner core for a constant value $\lambda_i$ of the diffusivity. At the outer boundary, $r_o$, the magnetic field is matched to a potential field in the exterior. In this study we systematically vary the strength of the applied dipole, $B_{dip}$.\\
\begin{table}
\caption{Simulations carried out with critical varied parameters given. The reference case is in bold. Hydrodynamic models H1 and H2 have stress-free and rigid lower boundary conditions respectively. \label{tab:Cases}}
\centering
\begin{tabular}{ l c c c c c c}
    \hline
    Simulation & $\rho_i$ & $B_{dip}$ & $1/d_\lambda$ & $\sigma_i$ & $\Lambda(0.8r_o)$ & Symbol \\
    \hline
    H1 & 1 & - & - & - & - & - \\
    H2 & 1 & - & - & - & - & - \\
    \hline
    B1.0 & 1 & 0.25 & $\ln(10^8)$ & 1 & $6.04\cdot 10^{-5}$ & \textcolor{blue}{$+$} \\
    \textbf{B1.1} & \textbf{1} & \textbf{0.5} & \textbf{ln(10}$^8$\textbf{)} & \textbf{1} & \textbf{2.42}$\cdot$\textbf{10}$^{-4}$ & \textcolor{blue}{$\times$} \\
    B1.2 & 1 & 1 & $\ln(10^8)$ & 1 & $9.67\cdot 10^{-4}$ & \textcolor{blue}{$\blacktriangleleft$} \\
    B1.3 & 1 & 2 & $\ln(10^8)$ & 1 & $3.87\cdot 10^{-3}$ & \textcolor{blue}{$\blacktriangleright$} \\
    \hline
    B2.0 & 1 & 0.5 & $\ln(10^8)$ & 0.25 & $6.04\cdot 10^{-5}$ & \textcolor{orange}{$+$} \\
    B2.2 & 1 & 0.5 & $\ln(10^8)$ & 4 & $9.67\cdot 10^{-4}$ & \textcolor{orange}{$\blacktriangleleft$} \\
    B2.3 & 1 & 0.5 & $\ln(10^8)$ & 16 & $3.87\cdot 10^{-3}$ & \textcolor{orange}{$\blacktriangleright$} \\
    \hline
    B3.0 & 1 & 0.5 & $\ln(10^6)$ & 0.25 & $2.80\cdot 10^{-4}$ & \textcolor{green}{$+$} \\
    B3.1 & 1 & 0.5 & $\ln(10^6)$ & 1 & $1.12\cdot 10^{-3}$ & \textcolor{green}{$\times$} \\
    B3.2 & 1 & 0.5 & $\ln(10^6)$ & 4 & $4.49\cdot 10^{-3}$ & \textcolor{green}{$\blacktriangleleft$} \\
    B3.3 & 1 & 0.5 & $\ln(10^6)$ & 16 & $1.80\cdot 10^{-2}$ & \textcolor{green}{$\blacktriangleright$} \\
    \hline
    B4.0 & 1 & 0.5 & $\ln(10^{10})$ & 0.25 & $1.30\cdot 10^{-5}$ & \textcolor{red}{$+$} \\
    B4.1 & 1 & 0.5 & $\ln(10^{10})$ & 1 & $5.21\cdot 10^{-4}$ & \textcolor{red}{$\times$} \\
    B4.2 & 1 & 0.5 & $\ln(10^{10})$ & 4 & $2.08\cdot 10^{-4}$ & \textcolor{red}{$\blacktriangleleft$} \\
    B4.3 & 1 & 0.5 & $\ln(10^{10})$ & 16 & $8.33\cdot 10^{-3}$ & \textcolor{red}{$\blacktriangleright$} \\
    \hline
    A1.0 & 4 & 0.25 & $\ln(10^8)$ & 1 & $1.58\cdot 10^{-5}$ & \textcolor{violet}{$+$} \\
    A1.1 & 4 & 0.5 & $\ln(10^8)$ & 1 & $6.34\cdot 10^{-5}$ & \textcolor{violet}{$\times$} \\
    A1.2 & 4 & 1 & $\ln(10^8)$ & 1 & $2.53\cdot 10^{-4}$ & \textcolor{violet}{$\blacktriangleleft$} \\
    A1.3 & 4 & 2 & $\ln(10^8)$ & 1 & $1.01\cdot 10^{-3}$ & \textcolor{violet}{$\blacktriangleright$} \\
    \hline
    A2.0 & 4 & 0.5 & $\ln(10^8)$ & 0.25 & $1.58\cdot 10^{-5}$ & \textcolor{brown}{$+$} \\
    A2.2 & 4 & 0.5 & $\ln(10^8)$ & 4 & $2.53\cdot 10^{-4}$ & \textcolor{brown}{$\blacktriangleleft$} \\
    A2.3 & 4 & 0.5 & $\ln(10^8)$ & 16 & $1.01\cdot 10^{-3}$ & \textcolor{brown}{$\blacktriangleright$}
\end{tabular}
\end{table}

\subsection{Numerical Methods}\label{sec:NumMethods}

All simulations in this study have been computed using the MHD code MagIC (available at https://github.com/magic-sph/magic). We use both the original Boussinesq version (see Wicht, 2002) and that which uses the anelastic approximation \cite{Jones_2011}. The governing equations given in section~\ref{sec:MHDequations} are solved by expanding both velocity (or \( \tilde{\rho} \)\textbf{u} in the anelastic cases) and magnetic fields into poloidal and toroidal potentials. For further details see \citeA{Christensen_2015}. The potentials are expanded in Chebychev polynomials in the radial direction and spherical harmonics up to a degree $\ell_{max}$ in the angular direction.\\
We use 145 radial grid-points for all simulations in the study. We use a \change{non-linear}{nonlinear} mapping function \cite{Tilgner_1999} to concentrate the grid-points around the transition from convecting to sub-adiabatic. This ensures both the boundary between the two layers as well as the shell boundary regions are well-resolved, as illustrated in Figure~\ref{fig:RadDist}. See the Appendix for details on the mapping.\\
For the reference case, labelled B1.1, we carried out one simulation without any imposed azimuthal symmetry, using azimuthal resolution $n_\phi=1280$ and without \change{hyper-diffusivity}{hyperdiffusivity}. For the other cases we introduced a four-fold azimuthal symmetry, reduced the number of grid-points to $n_\phi=864$ and applied \change{hyper-diffusion}{hyperdiffusion}, where the diffusion parameters (thermal and viscous) are multiplied by the factor
\begin{equation}
    \nu(\ell) = \kappa(\ell) = 1+D\left[ \frac{\ell+1-\ell_{hd}}{\ell_{max}+1-\ell_{hd}} \right]^\beta,
\end{equation}
for $\ell\geq\ell_{hd}$, where $\ell_{hd}=250$, $D=4$ and $\beta=2$. We verified that in the reference case the zonal winds formed and other features vital for our analysis did not change with imposed symmetry and \change{hyper-diffusion}{hyperdiffusion}.\\
All analysis was then based on the final stage of the simulations, which were integrated for 800,000 time-steps after they were fully equilibrated which is around $0.2\tau_\nu$ ($\sim20,000$ rotations). Equilibration required several viscous diffusion times for the first reference case. The other simulations in this study were started from the reference model as the initial state. It typically took one viscous diffusion time for the kinetic and magnetic energies to reach a statistical steady state with the modified parameters.

\section{Results}\label{sec:Results}

In our study we vary the strength of the imposed dipole field, $B_{dip}$, the electrical conductivity at the inner boundary, $\sigma_i$, and the conductivity scale height, $d_\sigma$. The parameters are summarised in Table~\ref{tab:Cases}. We explore the surface zonal wind profiles, their extension into the interior and the mechanisms by which they are quenched.\\

\subsection{Zonal Wind Distribution}\label{sec:UphiDist}

\begin{figure}
\noindent\includegraphics[width=14cm]{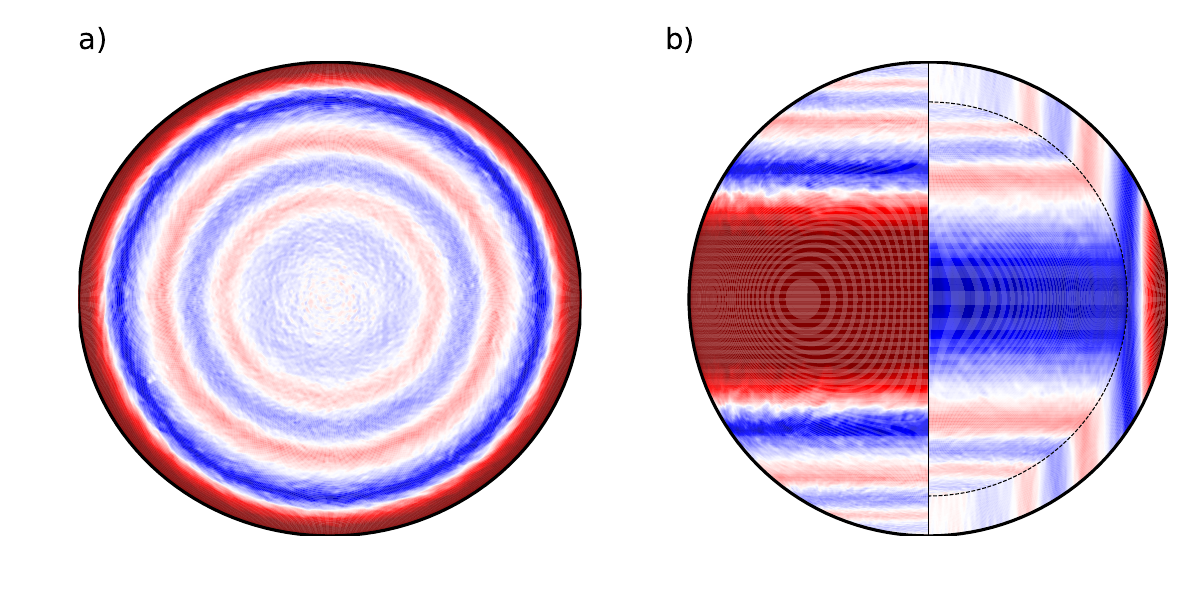}
\caption{A snapshot of the azimuthal flow, $u_\phi$, for the reference case B1.1. Both plots use the same colour-scale with a dynamic range of $\pm6000$, where red (blue) indicates prograde (retrograde) flow. a) View onto the surface of the spherical shell from the North Pole. b) Front view of the surface flow on the left and a cut down to the bottom of the convecting layer on the right.}
\label{fig:UphiSnap}
\end{figure}
\begin{figure}
\noindent\includegraphics[width=14cm]{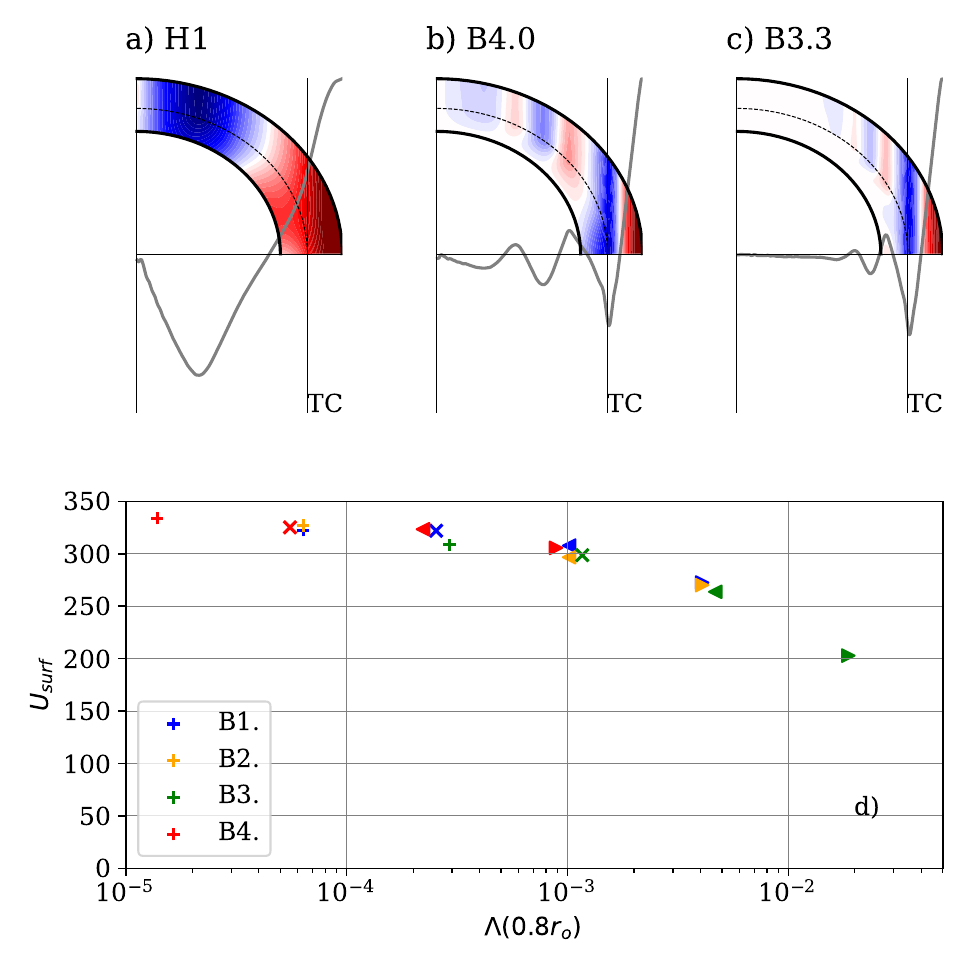}
\caption{Time-averaged axisymmetric zonal flow for simulations H1 in panel a), B4.0 in panel b) and B3.3 in panel c) (see Table \ref{tab:Cases}) with the same colour-scale as Figure~\ref{fig:UphiSnap}, with range $\pm6000$. On top of these are plotted the respective surface wind profiles as a function of $s$ for the hemisphere shown. The thin vertical lines indicate the locations of the tangent cylinders associated with the bottom of the convective region, TC. d) shows the average zonal flow velocity inside the TC (defined by eq. \ref{eq:Usurf}), as a function of the local Elsasser number evaluated at $0.8r_o$. See table~\ref{tab:Cases} for the symbols for each case.}
\label{fig:UphiDist}
\end{figure}

The snapshot of our reference case B1.1, in Figure~\ref{fig:UphiSnap}, shows that these simulations reproduce one of the key features found in the measurements of the zonal flows of the two gas giants: a set of alternating zonal jets reaching up the high latitudes. The equatorial prograde jet and its flanking retrograde jets dominate, but slightly weaker flows also persist up to the poles. These extend geostrophically, i.e. invariant with respect to $z$ which is parallel to the rotation axis, throughout the convective region. We show the time-averaged, axisymmetric zonal flow for only one hemisphere of the hydrodynamic comparison case H in Figure~\ref{fig:UphiDist}a. Plotted on top of this is the surface profile as a function of the cylindrical coordinate $s=r\sin\theta$, i.e. the distance from the axis of rotation. We observe that in case H1, without either the additional magnetic forces or a mechanical rigid boundary condition which can act as a proxy for some force that brakes the jets, the jets are much wider and their amplitude (in this case that of the only retrograde jet present inside the tangent cylinder) only decreases slightly when reaching the stable layer. We note that a similar purely hydrodynamic simulation with a stress-free flow boundary shown in Figure 3d of \citeA{Wulff_2022} also shows a zonal flow pattern unlike that of Jupiter or Saturn, with a few strong jets inside the tangent cylinder (TC) that decay only weakly towards the inner boundary. The differences to the present case can be attributed to the anelastic approximation and a larger degree of stability in \citeA{Wulff_2022}.\\
However, Figures~\ref{fig:UphiDist}b and c demonstrate that under the influence of finite conductivity and a large-scale magnetic field, the zonal flows develop a multiple jet structure. Furthermore, the jets are quenched effectively in the stable layer. The two cases shown, B4.0 and B3.3, are the extremes in the study. In B4.0 the conductivity starts out rather small at the inner boundary and drops rapidly with radius. This case shows strong zonal winds in the tangent cylinder reaching the polar region. In B3.3 the conductivity at $r_i$ starts out rather large and drops more weakly with radius. Here, significant jets are still found at mid-latitude, but they fade out at the high latitudes. The vertical extent of the convective region, i.e. the depth of the stable layer, is not altered in the study so the TC is in the same location and the equatorial prograde jet has the same width, with the peaks of the flanking retrograde jets being located on the TC.\\
The relation between the jet widths and jet amplitudes was confirmed to obey Rhines scaling well, when taking the convective region as the shell thickness (following the methodology detailed in \citeA{Gastine_2014}). This predicts that narrower jets are also weaker.
In order to quantify the strength of the axisymmetric zonal flow inside the tangent cylinder (TC), we define the average surface zonal flow amplitude in this region as:
\begin{equation}\label{eq:Usurf}
    U_{surf} = \frac{1}{2\theta_c}\left(\int_0^{\theta_c} \sqrt{\langle\overline{u}_\phi(r_o, \theta)\rangle^2}\sin\theta \, d\theta + \int_{\pi-\theta_c}^\pi \sqrt{\langle\overline{u}_\phi(r_o, \theta)\rangle^2}\sin\theta \, d\theta \right),
\end{equation}
where $\sqrt{\langle\overline{u}_\phi(r_o, \theta)\rangle^2}$ is the time-averaged, axisymmetric, rms surface zonal flow and $\theta_c=\sin^{-1}(r_c/r_o)$, i.e. the colatitude associated with the location of the TC at the surface. This definition broadly captures both the extent and strength of the zonal flow and facilitates a comparison between all cases.\\
We observe that simulations with a stronger imposed dipole field strength, $B_{dip}$, and those with higher electrical conductivity, $\sigma$, have weaker winds inside the TC. We use a local Elsasser number:
\begin{equation}
    \Lambda(r)=\frac{B_{dip}(r)^2\sigma(r)}{\tilde{\rho}(r)\Omega},
\end{equation}
as a proxy for the local strength of the Lorentz force, relative to the Coriolis force, although in this study we only explicitly test the dependency on $B_{dip}$ and $\sigma$. This expresses not only the radial variation of the electrical conductivity but also the $r^{-3}$ dependence of the dipole field strength (in our definition we use the axial dipole field amplitude at the poles).\\
In Figure~\ref{fig:UphiDist}d we therefore plot $U_{surf}$, as a function of the Elsasser number evaluated at $0.8r_o$, i.e. in the upper part of the stable layer, just below $r_s$. The extremes of our parameter sweep are $\Lambda(0.8r_o)=1.30\cdot 10^{-5}$ in case B4.0, up to $1.80\cdot 10^{-2}$ in case B3.3 (see Table~\ref{tab:Cases}).\\
The plot suggests that if magnetic forces remain insignificant near the SSL boundary, strong zonal winds can develop and be maintained in the overlying convecting region and are independent of the magnetic effects coming into play deeper in the stable region. However, when magnetic effects become more pronounced in the upper part of the stable layer, the zonal flow inside the TC becomes somewhat more diminished, in particular at high latitudes. Within our parameter sweep this is not a dramatic effect. As our focus is on models that have strong jets inside the TC so we do not go beyond case B3.3. We would expect these to disappear if the semiconducting region begins at even shallower depths and $\Lambda(0.8r_o)$ is increased by even just one more order of magnitude.\\

\subsection{Flow Amplitude Versus Depth}\label{sec:Track}

\begin{figure}
\noindent\includegraphics[width=14cm]{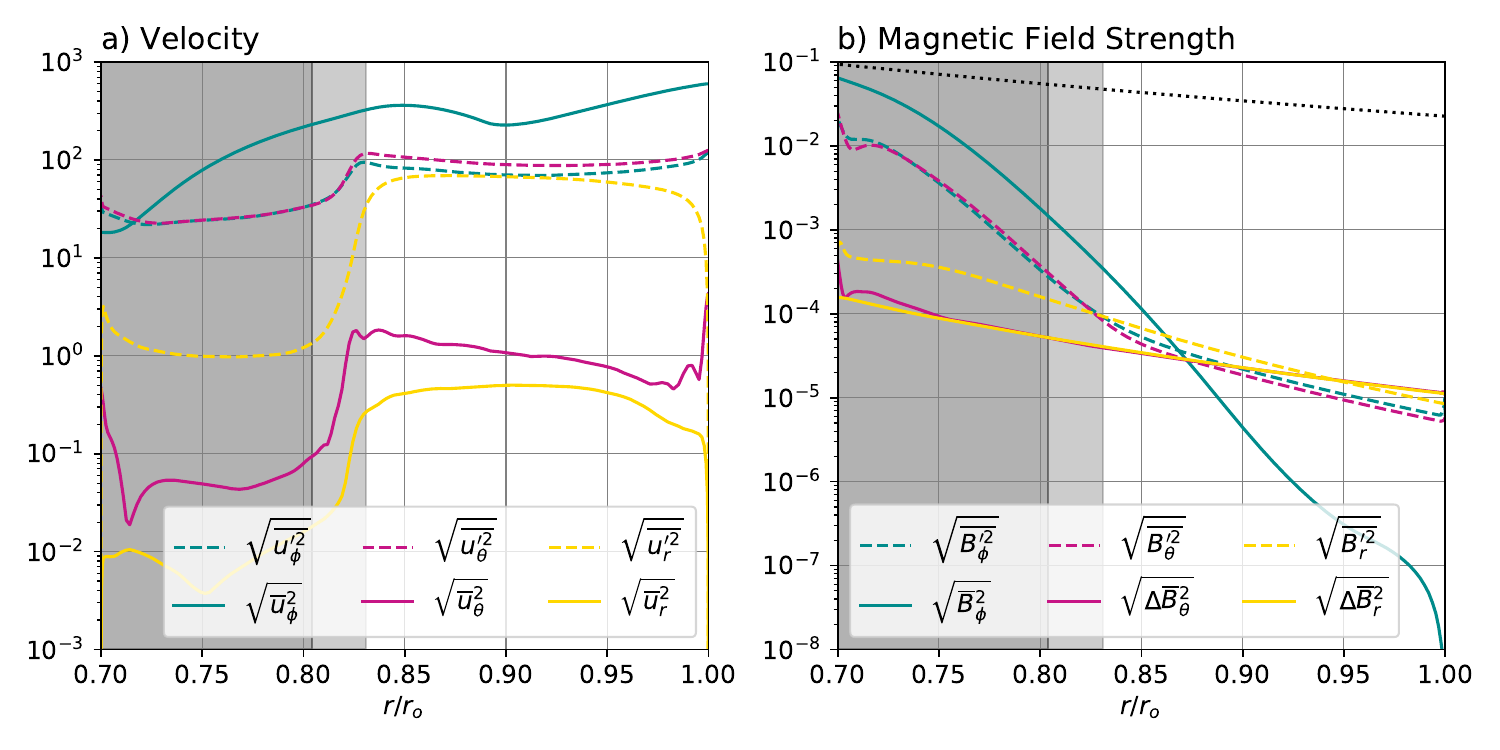}
\caption{Radial profiles of time- and horizontally-averaged a) velocity and b) magnetic field strength (given in $\Lambda$) for the reference case B1.1. The dashed lines show the average non-axisymmetric flow (field strength) and the solid lines are the axisymmetric parts, where colours indicate the three components. For $\overline{B}_\theta$ and $\overline{B}_r$ we subtract the dipole component, of which the average amplitude is shown by the black dotted line. The dark grey (grey) shading indicates the (transition into the) SSL.}
\label{fig:UB_avg}
\end{figure}

Figure~\ref{fig:UB_avg}a shows the horizontally averaged rms velocity components for the reference simulation as a function of radius, where solid lines show the axisymmetric components (labelled with an overbar) and dashed lines the non-axisymmetric components (indicated by a prime). In the convective region, the convective flow amplitude ($u^\prime_\phi$, $u^\prime_\theta$ and $u^\prime_r$) is almost an order of magnitude weaker than the rms zonal wind amplitude (the jet peaks themselves are even stronger). Upon reaching the SSL, radial flow components are quenched most effectively and amplitudes drop by almost two orders of magnitude. At least part of the remaining radial motion seen in figure~\ref{fig:UB_avg} may represent wave motion (gravity waves, inertial waves) and no overturning motion. Right at the SSL boundary both the latitudinal component of the meridional flow, $\overline{u}_\theta$, and the horizontal components of the convective flow, $u_\phi^\prime$ and $u_\theta^\prime$, increase very slightly which may be attributed to the deflection of the radial flows. However, further into the SSL all other flow components are damped. We analyse this in more detail for the zonal flow.\\
\begin{figure}
\noindent\includegraphics[width=14cm]{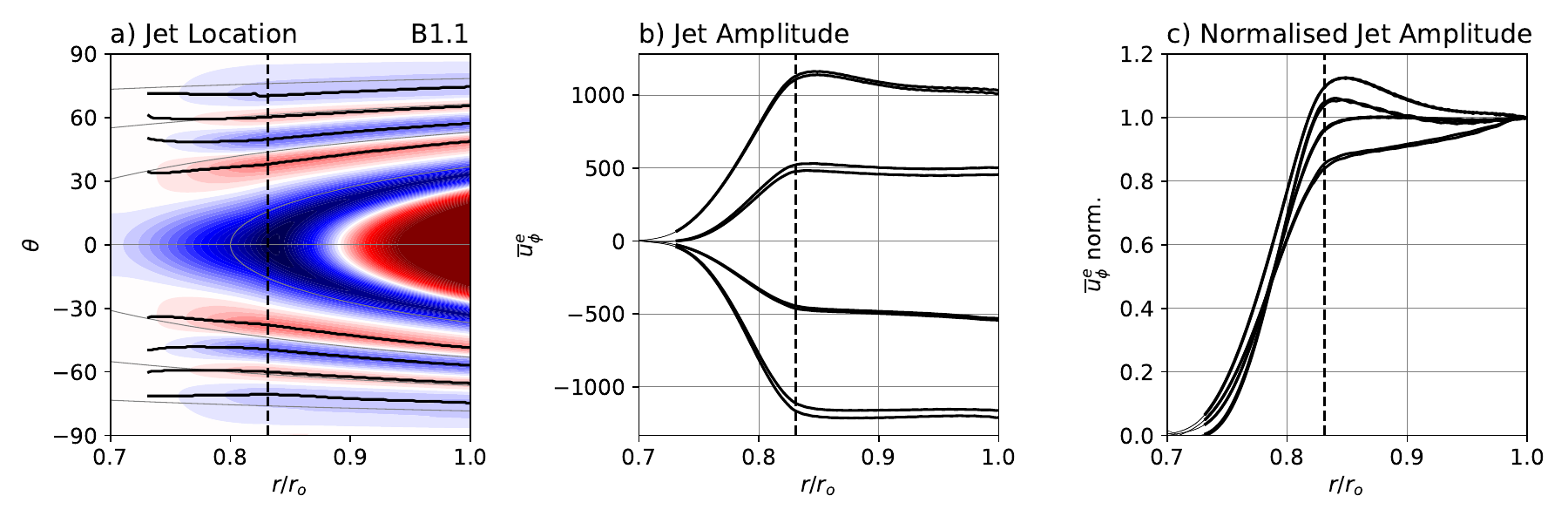}
\caption{Illustration of the jet tracking method described, applied to the reference case B1.1. a) shows the zonal flow pattern. The black lines show the locations of the zonal flow extrema (denoted by superscript $e$) and the grey lines indicate lines of constant $s$. b) shows the peak amplitudes of these 8 jets as a function of radius and in c) we normalise this by the jet flow velocity at $r=r_o$. The location of $r_c$ is indicated by a vertical dashed line in all three plots.}
\label{fig:TrackJets}
\end{figure}
We track the jet amplitude as a function of radius in the SSL. This is illustrated in Figure~\ref{fig:TrackJets}a where we show the locations of the maxima/minima of the jets inside the TC for simulation B1.1. This tracking is vital as the locations of the peak velocity is no longer $z$-invariant in the SSL in contrast to the convective region, as can be seen in figure~\ref{fig:UphiDist}b and c.\\
Figure~\ref{fig:TrackJets}b shows $\overline{u}_\phi^e$ along the centres of these jets (we use superscript $e$ to denote the extrema of $\overline{u}_\phi$ as a function of latitude). This also highlights the strong equatorial symmetry of these particular simulations where the northern/southern hemisphere jet pairs have almost identical velocity profiles. Finally, Figure~\ref{fig:TrackJets}c shows the same velocity profiles, with each one normalised by the respective jet velocity at $r_o$. This plot clearly illustrates that the relative decay with depth is rather similar for all jets, independent of their location inside the TC.\\
\begin{figure}
\noindent\includegraphics[width=14cm]{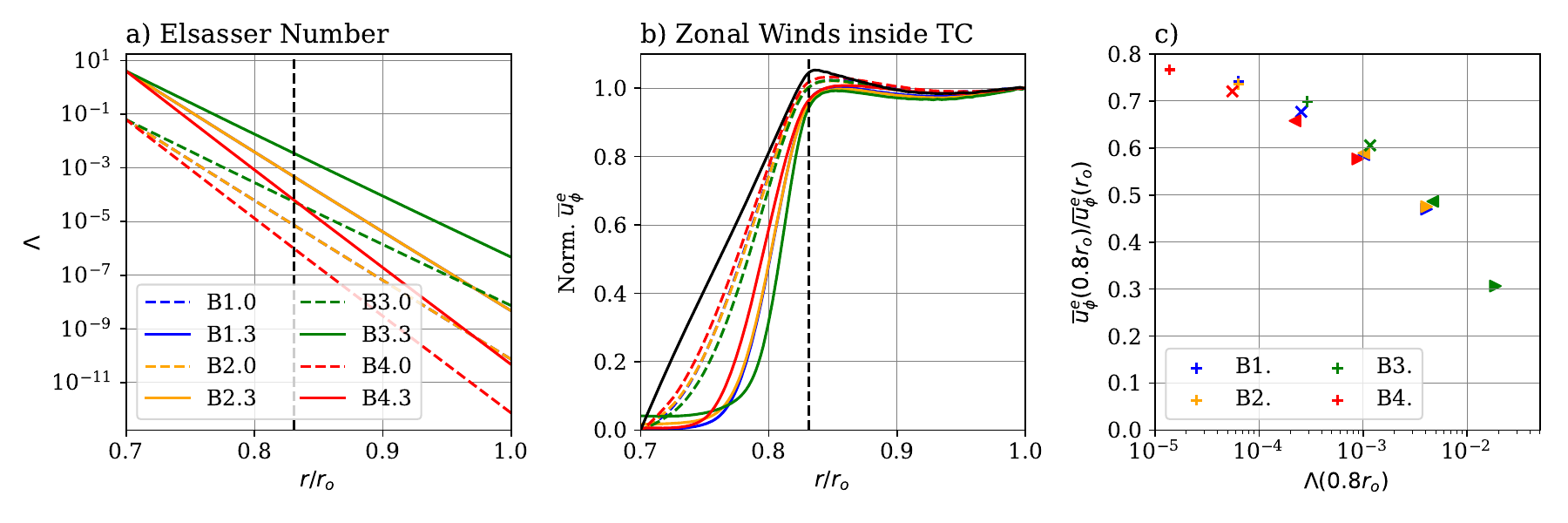}
\caption{a) Elsasser number as a function of depth for the end-members of each of the Boussinesq sets. b) shows the normalised jet amplitude profiles as shown in Figure~\ref{fig:TrackJets}c where a single profile is obtained for each case by averaging over all 8 jets. The black line corresponds to hydrodynamic comparison case H2, which has a rigid boundary condition. c) ratio of zonal flow amplitude at $0.8r_o$ and the jet flow velocity at the surface, obtained from the averaged profiles shown in b) and the remaining models omitted on this plot. See Table~\ref{tab:Cases} for the symbols for each case.}
\label{fig:Uphi_Prof_Els}
\end{figure}
We average the radial profiles of all jets inside the TC, normalised by their velocity at the bottom of the convecting region, for each case to quantify the zonal wind decay. Figure~\ref{fig:Uphi_Prof_Els}b compares averaged profiles for the end-member simulations of each of the Boussinesq sets while Figure~\ref{fig:Uphi_Prof_Els}a shows the respective Elsasser number profiles. In all cases with magnetic effects the velocity decreases more sharply in the stable layer than in the purely hydrodynamic case (black line in  figure~\ref{fig:Uphi_Prof_Els}b). In addition, the hydrodynamic case required a no-slip condition at the inner boundary to reach a small velocity near the bottom, whereas the magnetic cases assume stress-free. Profiles with identical $\Lambda(r)$ (sets B1. and B2.) nearly perfectly overlap which highlights that the Elsasser number is the crucial parameter here; doubling the axial dipole field strength has exactly the same effect as quadrupling the electrical conductivity. When considering the other profiles shown we clearly see that in simulations with the lowest Elsasser numbers the decay of the zonal wind is very gradual. This is illustrated in figure~\ref{fig:Uphi_Prof_Els}c where we plot the ratio of the jet amplitude at $0.8r_o$ and the amplitude at $r_o$, again averaging over all 8 jets to obtain one value per simulation. Therefore, magnetic effects are crucial in reducing the penetration distance of zonal winds into the SSL.

\subsection{Magnetic Field Induction}\label{sec:BInd}

Figure~\ref{fig:UB_avg}b shows the horizontally averaged induced magnetic field components for the reference case. The induced axisymmetric toroidal field is almost as strong as the dipole field at the lower boundary, for this case, but drops off rapidly with radius. The induced axisymmetric radial and latitudinal fields, $\Delta\overline{B}_r$ and $\Delta\overline{B}_\theta$, i.e. the perturbations of the imposed poloidal field, are almost three orders of magnitude smaller but do not drop off in amplitude as sharply over the SSL.\\
\begin{figure}
\noindent\includegraphics[width=14cm]{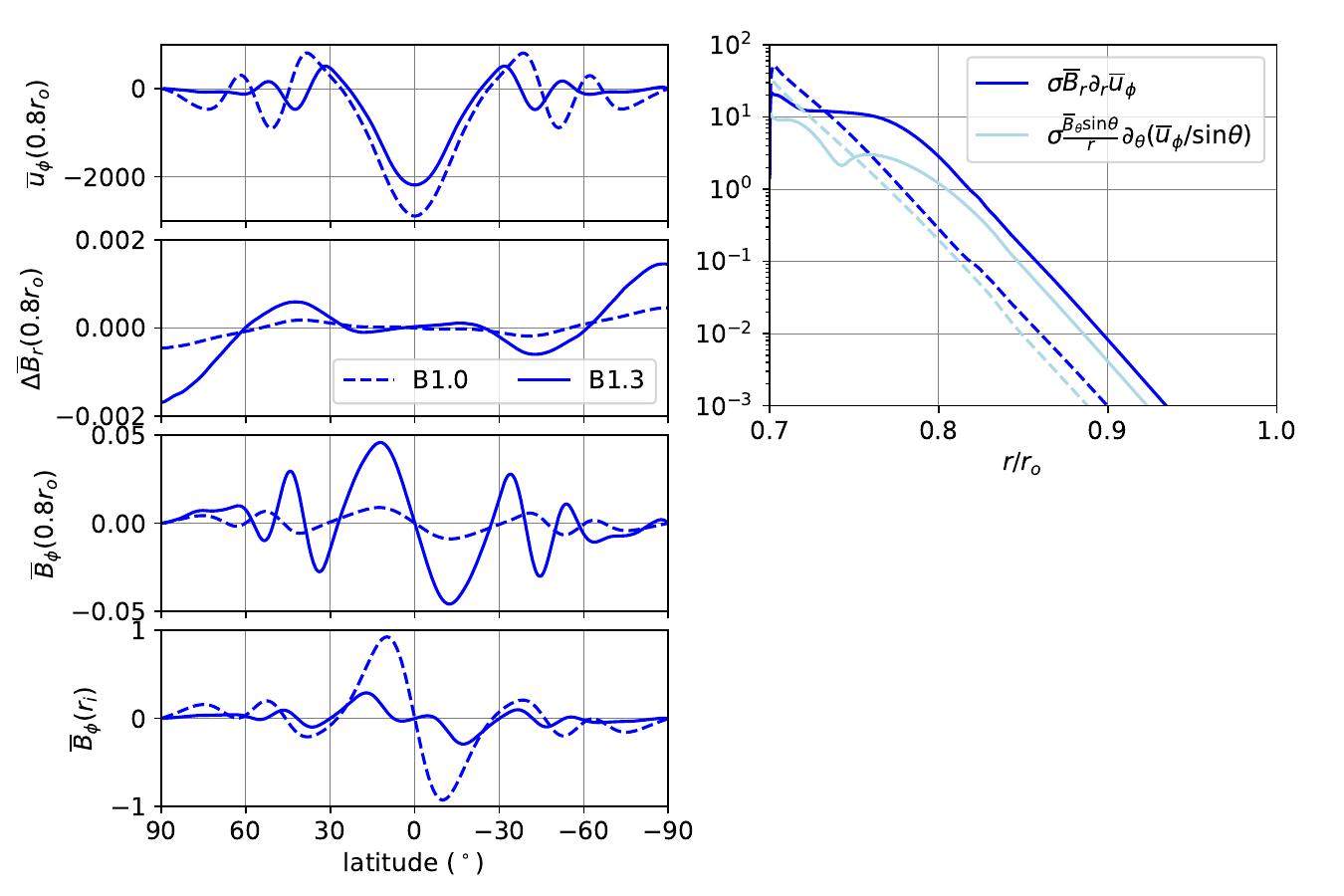}
\caption{Left: Latitudinal profiles for cases B1.0 and B1.3. The top panels show $\overline{u}_\phi$ and $\Delta\overline{B}_r=\overline{B}_r-\overline{B}_r^{dip}$ on $0.8r_o$. The lower panels are $\overline{B}_\phi$ at $0.8r_o$ and $r_i$. Right: Horizontally averaged, rms amplitude of the two terms in the $\Omega$-effect for the same cases, where dashed(solid) corresponds to B1.0(B1.3) and dark(light) blue corresponds to the radial(latitudinal) shear term.}
\label{fig:B_Ind_comp}
\end{figure}
We investigate the difference in the induction when changing the imposed axial dipole field strength in Figure~\ref{fig:B_Ind_comp}. On the left the top three plots are latitudinal profiles of $\overline{u}_\phi$, $\Delta\overline{B}_r=\overline{B}_r-\overline{B}_r^{dip}$ and $\overline{B}_\phi$ at $0.8r_o$. These are all as we may expect, with B1.3 having the strongest induced magnetic fields. This is due to it having the largest imposed dipole amplitude, eight times stronger than case B1.0 with the same conductivity profile, leading to a larger $\Omega$-effect. The $\Omega$-effect describes the induction of axisymmetric toroidal field by the shearing of the axisymmetric poloidal field by differential rotation and has the two components $\overline{B}_r \partial_r\left(\frac{\overline{u}_\phi}{r}\right)$ and  $\frac{\overline{B}_\theta\sin\theta}{r}\partial_\theta\left(\frac{\overline{u}_\phi}{\sin\theta}\right)$. From this we see that the assumption that a stronger dipole will lead to a stronger induced field holds only when the simulations with different $B_{dip}$ have similar distributions of $\overline{u}_\phi$. However, in  Figure~\ref{fig:Uphi_Prof_Els}\textbf{b} we see that the zonal wind is quenched very effectively in case B1.3, so $\overline{u}_\phi$ is almost zero at mid-depth of the stable layer, while this only happens near the bottom in case B1.0. Therefore, the $\Omega$-effect is stronger for case B1.0 than for B1.3 in the lower third of the stable layer and the amplitude of the induced field, $\overline{B}_\phi$, exceeds that of B1.3 significantly, as can be seen in the bottom panel of Figure~\ref{fig:B_Ind_comp}. In fact, for case B1.0 it also exceeds the amplitude of the imposed dipole field at the lower boundary which for this case is $B_r(r_i, \theta=0)=-0.25$. Thus, the model, B1.0, which has the most interaction between the zonal flow and the electrically conducting region (i.e. over the greatest depth range) actually has the weakest induced field strength near the top of its conducting region.\\
The morphology of the induced $\overline{B}_\phi$ field can also be better understood by comparing the contributions of the two terms that make up the $\Omega$-effect; the radial and the latitudinal shear of the zonal wind. This is shown on the right in Figure~\ref{fig:B_Ind_comp}, where the rms amplitude of the two terms has been averaged horizontally to produce radial profiles. For both cases, B1.0 and B1.3, the radial shear is the more dominant term throughout the shell. Therefore, the decay of the jets with depth produces a stronger gradient than the transition between oppositely flowing jets. This also leads to the induced azimuthal field being strongest almost exactly on the zonal wind peaks.

\subsection{Zonal Wind Truncation Mechanism}\label{sec:Trunc}

\begin{figure}
\noindent\includegraphics[width=14cm]{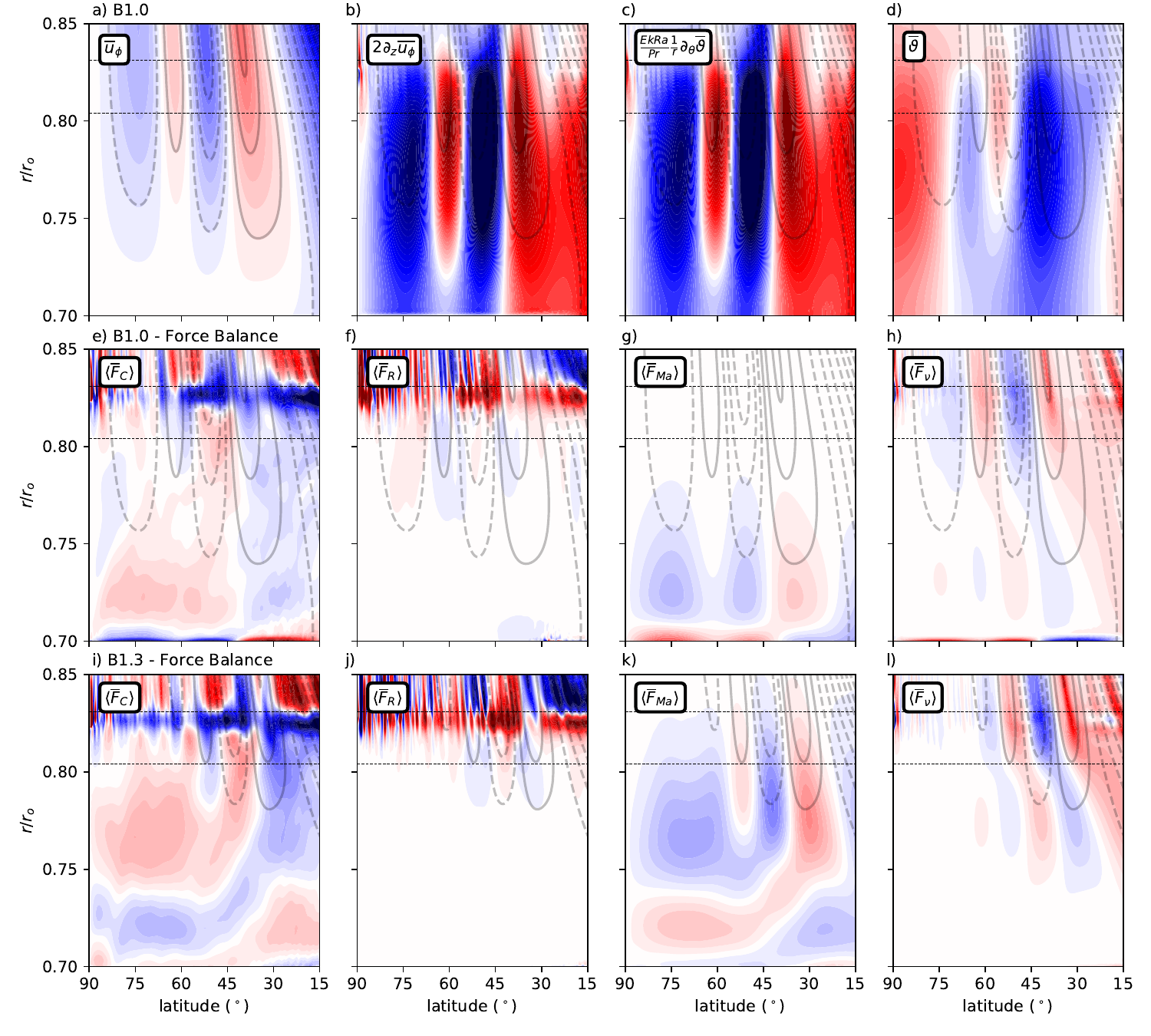}
\caption{The transition region and SSL, shown for the northern hemisphere. All terms are zonally and temporally averaged. The top row shows, for case B1.0: a) $\overline{u}_\phi$, the zonal flow, b) the vertical gradient of the zonal flow, c) the latitudinal temperature fluctuation and d) the temperature. The thin horizontal lines indicate $r_c$ and $r_s$, i.e. the top and bottom of the transition region into the SSL. Lower rows: azimuthal component of the force balance, shown for cases e)-h) B1.0 and i)-l) B1.3. The panels show e), i) Coriolis term $\overline{F}_C$; f), j) Reynolds stresses $\overline{F}_R$; g), k) Lorentz forces $\overline{F}_{Ma}$; and h), l) viscosity $\overline{F}_\nu$. Solid (dashed) grey contours indicate positive (negative) $\overline{u}_\phi$.}
\label{fig:Meri}
\end{figure}
A thermo-magnetic wind equation can be derived by taking the $\phi$-component of the curl of the Navier-Stokes equation, then averaging over azimuth and assuming steady state:
\begin{eqnarray}
    0 & = & \frac{2E}{s}\overline{u}_\phi\partial_z\overline{u}_\phi - Es\overline{u}_s\partial_s\frac{\overline{\omega}_\phi}{s}-E\overline{u}_z\partial_z\overline{\omega}_\phi +2\partial_z\overline{u}_\phi  -\frac{Ra E}{Pr}\frac{1}{r}\partial_\theta \overline{\vartheta} \nonumber \\
    && + \frac{1}{Pm}\overline{\left[\nabla\times\left( \mathbf{j}\times\mathbf{B}\right)\right]}_\phi+ E\overline{\left[\nabla^2\mathbf{\omega}\right]}_\phi,
\end{eqnarray}
where $\mathbf{\omega}=\nabla\times\mathbf{u}$ is the vorticity and $\mathbf{j}=\nabla\times\mathbf{B}$. The first three terms are from the advection term, the fourth and fifth terms are from the Coriolis force and buoyancy, respectively. The last two terms are from the Lorentz force and the viscous force. We find that in our simulations, the equation can be reduced to:
\begin{equation}
    2\partial_z\overline{u}_\phi \approx \frac{Ra E}{Pr}\frac{1}{r}\partial_\theta \overline{\vartheta}, \label{eq:MTW}
\end{equation}
as all other terms were found to be negligible. This is shown in Figure~\ref{fig:Meri}b and c where we plot the vertical gradient of the zonal wind (first term of eq.~\ref{eq:MTW}) and the latitudinal temperature gradient (second term of eq.~\ref{eq:MTW}). The two are in nearly perfect balance; the magnetic term of the thermo-magnetic wind equation is negligibly small. As in the hydrodynamic simulations in \citeA{Wulff_2022}, the decrease of the zonal wind in the stable layer is controlled by a thermal wind balance. The associated density perturbation is caused by a meridional flow. As Lorentz forces play a critical role for the penetration of the winds into the stable layer, this should happen via their influence on the meridional circulation. To elucidate this, we consider the time-averaged (denoted by $\langle \rangle$) axisymmetric, azimuthal component of the Navier-Stokes equation, given by:
\begin{eqnarray}
    0 =& \langle\overline{F}_{Ad}\rangle + \langle\overline{F}_C\rangle + \langle\overline{F}_R\rangle + \langle\overline{F}_\nu\rangle + \langle\overline{F}_{Ma}\rangle + \langle\overline{F}_{Mna}\rangle\, ;\nonumber\\
    \overline{F}_{Ad} =& \frac{\overline{u}_s}{s}\partial_s(s\overline{u}_\phi) + \overline{u}_z\partial_z(\overline{u}_\phi) \nonumber\\
    \overline{F}_C =& \frac{2}{E} \overline{u}_s \nonumber\\
    \overline{F}_R = & \frac{1}{s^2}\partial_s\left[s^2 \overline{u'_s u'_\phi}\right] + \partial_z \left[ \overline{u'_z u'_\phi}\right] \nonumber\\
    \overline{F}_\nu = & - \frac{1}{s^2} \partial_s \left[s^3 \partial_s\left(\frac{\overline{u}_\phi}{s}\right)\right] - \partial_z \left[ \partial_z\left(\overline{u}_\phi\right)\right] \nonumber\\
    \overline{F}_{Ma} = & \frac{-1}{E Pm}\left[\frac{1}{s^2}\partial_s\left(s^2\overline{B_\phi}\,\overline{B_s}\right) +\partial_z\left(\overline{B_\phi}\,\overline{B_z}\right)\right] \nonumber\\
    \overline{F}_{Mna} = & \frac{-1}{E Pm}\left[ \frac{1}{s^2}\partial_s\left(s^2\overline{B_\phi^\prime B_s^\prime}\right) +\partial_z\left(\overline{B_\phi^\prime B_z^\prime}\right)\right]. \label{eq:ZonForce_cyl}
\end{eqnarray}
This includes the: `advective' force $\overline{F}_{Ad}$, Coriolis force $\overline{F}_C$ and viscous force $\overline{F}_\nu$. The forces associated with the Reynolds stresses and the Maxwell stresses are $\overline{F}_R$, and $\overline{F}_{Ma}$ and $\overline{F}_{Mna}$ respectively, where the former is the contribution from the large-scale (axisymmetric) magnetic field components and the latter is from the small-scale (non-axisymmetric) field components.\\
We find that the advective force remains negligibly small and therefore omit it in Figure~\ref{fig:Meri} where we show the zonal force balance. Furthermore, the Maxwell stresses arising from the correlation of the small-scale magnetic field components, $\overline{F}_{Mna}$, also remain very small, even at depth and are thus also not shown in Figure~\ref{fig:Meri}. This is because in our study the stable layer suppresses small-scale flows so effectively. The conductivity distribution implies that the Lorentz forces only start acting in the SSL in these simulations, where only very weak non-axisymmetric induced magnetic field components contribute.\\
Figures~\ref{fig:Meri}e and i show the Coriolis force, which is directly proportional to the $s$-component of the meridional flow. This meridional flow is driven in the convecting and transition regions, where the associated Coriolis force is balanced mainly by the Reynolds stresses. The Reynolds stress force is enhanced in the transition region, by the same mechanism as in the purely hydrodynamic study \cite{Wulff_2022}, where radial flows and also all small-scale motion is quenched (see Figure~\ref{fig:UB_avg}). Therefore there is a sharp drop-off in the correlation of the convective flows just below $r_c$, leading to large derivatives with respect to $s$ and $z$ (see eq.~\ref{eq:ZonForce_cyl} for the definition of $\overline{F}_R$). The large Reynolds stress force is primarily balanced by the Coriolis force $\overline{F}_C$ of an enhanced meridional circulation. Inside the SSL there is a good match of $\overline{F}_C$ and the force associated with the Maxwell stresses, $\overline{F}_{Ma}$ (Figures~\ref{fig:Meri}g and k). This is the essential difference to the hydrodynamic models where only viscosity can balance $\overline{F}_C$ in this region. In the MHD case, the meridional flow remains significant in the SSL, so temperature (entropy) perturbations are induced (see Figure~\ref{fig:Meri}d) and the zonal flow can be quenched more effectively in the SSL. For a retrograde jet ($\overline{u}_\phi<0$) in the northern hemisphere the azimuthal component of the (divergence of) Reynolds stress which drives the jet is generally positive in the convective layer. It is balanced by the negative Coriolis force associated with the meridional circulation that is generally poleward in the upper layer in such a jet. Mass conservation requires that the meridional flow points in $-z$-direction on the poleward side of the jet and in $+z$-direction at the equatorward side. The circulation closes in the stable layer, where it points in $+s$-direction. Here the associated Coriolis force is balanced by Lorentz forces, or purely by viscous forces in the absence of a magnetic field. In the stable layer the sinking and rising branches of the meridional flow advect temperature (entropy), such that a positive anomaly is created on the poleward side of the jet and a negative anomaly on the equatorward side, seen in Figure~\ref{fig:Meri}d. Hence $\partial\vartheta / \partial \theta$ is negative in an eastward jet in the stable layer and according to eq.~\ref{eq:MTW} the jet velocity must decrease with z, or in other words the amplitude of the retrograde jet must drop with depth.\\
This is broadly in agreement with the mechanism proposed by \citeA{Christensen_2020}, where the winds were driven by an ad-hoc force rather than self-consistently by Reynolds stresses. We observe that the viscous force also plays a significant role in the SSL, as the zonal flow velocity is decreasing rapidly. At the much lower Ekman numbers that apply to the gas planets, viscosity is expected to play no significant role.\\
Comparing the force balances of two models we first note that in case B1.3 where the dipole strength is increased (Figure~\ref{fig:Meri}i-l), the Lorentz forces already begin to act in the SSL transition region. This illustrates how they are able to impact the structure of the zonal winds in the top part of the stable region  and impede jet formation at mid-high latitudes. Deeper in the stable region similar meridional circulation cells develop for both models, to balance the Lorentz forces. However, they are shifted upwards in case B1.3 relative to B1.0. In model B1.0 the winds only reach near-zero amplitude near the inner shell boundary and the transition from equator-ward (pole-ward) flow in the high- (mid-) latitude region to oppositely flowing meridional circulation occurs close to this boundary. In case B1.3 the winds are already quenched at around $0.74r_o$ which is where the circulation cells are centred in this model.

\subsection{Anelastic Simulations}

\begin{figure}
\noindent\includegraphics[width=14cm]{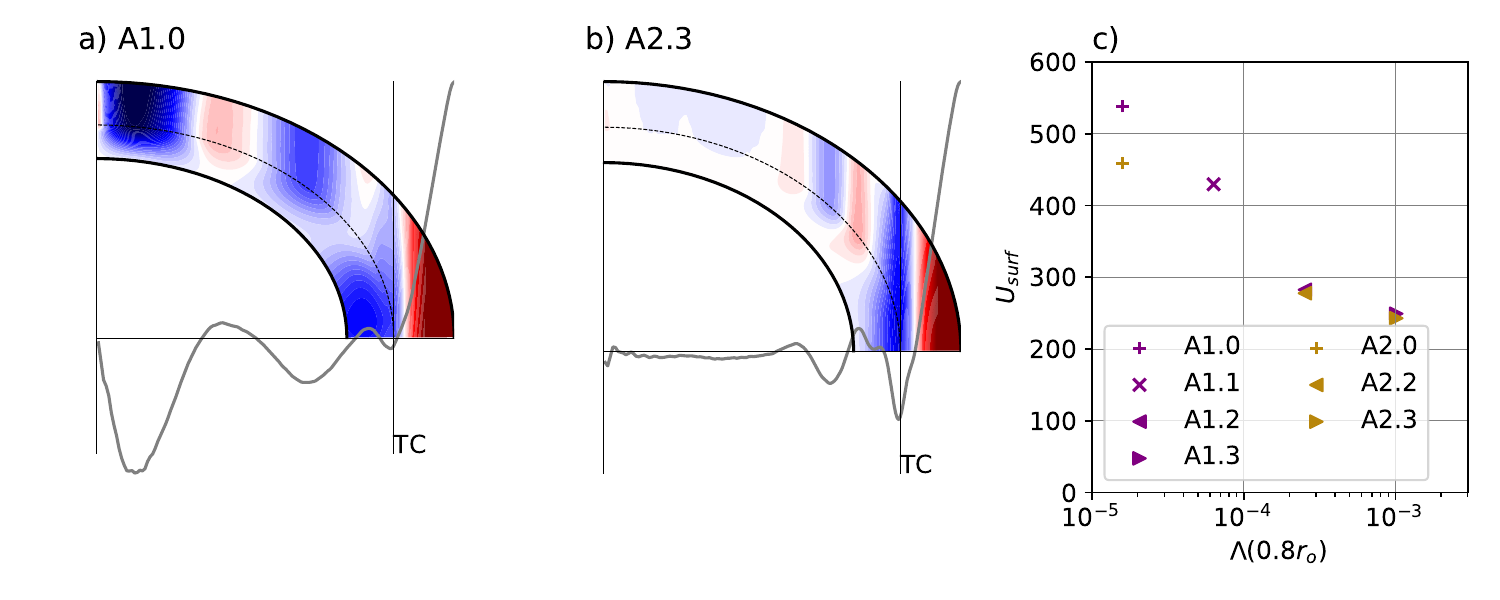}
\caption{a) and b) show the time-averaged axisymmetric zonal flow for simulations A1.0 and A2.3 (see Table \ref{tab:Cases}) with the same colour-scale as Figure~\ref{fig:UphiSnap}, with range $\pm6000$. On top of these are plotted the respective surface profiles as a function of $s$ for the hemisphere shown. The thin vertical lines indicate the locations of the tangent cylinders associated with the bottom of the convective region, TC. \textbf{c)} shows the average zonal flow velocity inside the TC (defined by eq. \ref{eq:Usurf}), as a function of the local Elsasser number evaluated at $0.8r_o$. See table~\ref{tab:Cases} for the symbols for each case.}
\label{fig:UphiDist_A}
\end{figure}
For seven models with different field strengths and conductivity profiles we replaced the Boussinesq approximation by the anelastic approximation in order to test its impact on the results (sets A1. and A2. in Table~\ref{tab:Cases}). Qualitatively, the zonal flows formed in these simulations are very similar to their Boussinesq counterparts, with the strongest jets being the prograde equatorial jet and its flanking retrograde jets, complemented by another four weaker jets inside the tangent cylinder (see figures~\ref{fig:UphiDist_A}a and b). In these simulations we also observed some time-variability in the zonal flow structure, similar to that discussed in \citeA{Wulff_2022}, which we do not explore further within this work.\\
Figure~\ref{fig:UphiDist_A} is the counterpart to Figure~\ref{fig:UphiDist}, showing the dependence of the rms zonal flow amplitude at the surface, inside the TC, on the local Elsasser number. While $\Lambda$ covers a smaller range than the Boussinesq study, the same trend is clearly seen: stronger winds develop in models where magnetic effects, characterised by $B_{dip}^2\sigma$, become significant only deeper into the stable layer.\\
\begin{figure}
\noindent\includegraphics[width=14cm]{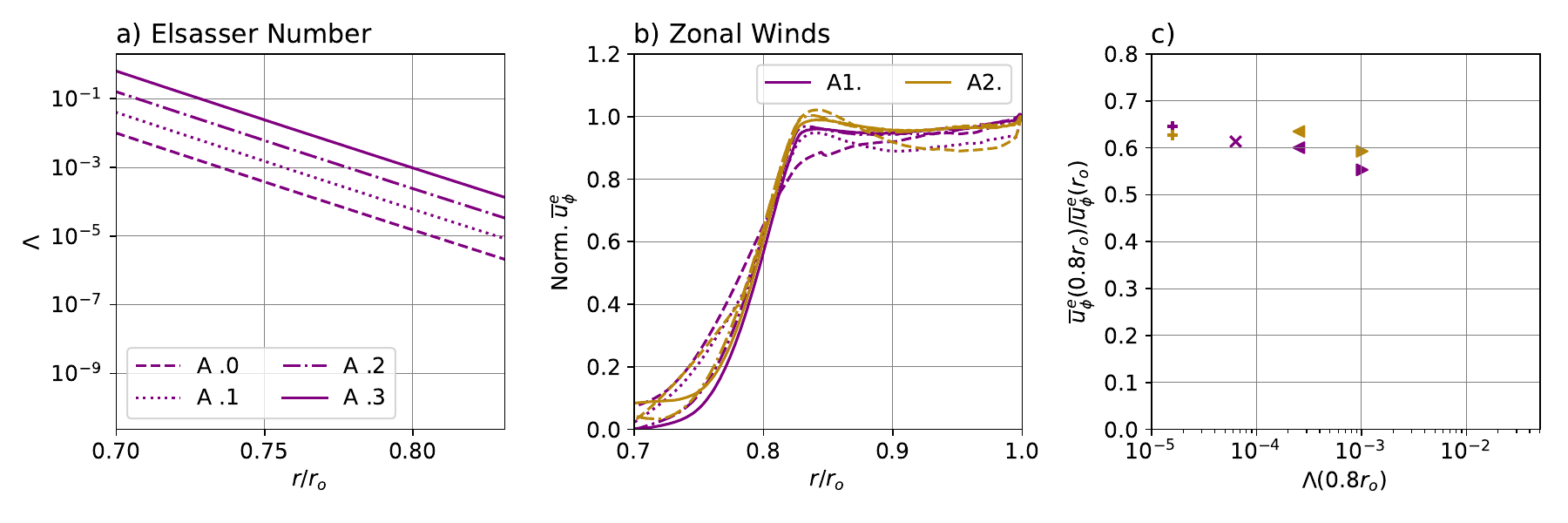}
\caption{a) Elsasser number as a function of depth for all 7 anelastic models. b) shows the normalised jet amplitude profiles as shown in figure~\ref{fig:TrackJets}c where a single profile is obtained for each case by averaging over all 8 jets. c) shows ratio of zonal flow amplitude at $0.8r_o$ and the jet flow velocity at $r=r_o$, obtained from the averaged profiles shown in b). See Table~\ref{tab:Cases} for the symbols for each case.}
\label{fig:Uphi_Prof_Els_A}
\end{figure}
We also test the relationship of the local Elsasser number and the zonal wind penetration distance for these anelastic models. We use the same analytical technique described in section~\ref{sec:Track} and track the jet amplitudes in the stable layer. This is shown in Figure~\ref{fig:Uphi_Prof_Els_A}. As Figure~\ref{fig:Uphi_Prof_Els_A}a shows, the two sets A1. and A2. have the same 4 radially varying Elsasser number profiles (with A1.1 forming part of both sets). However, in one set the axial dipole field strength, $B_{dip}$, is varied while in the other the electrical conductivity profile is varied (the conductivity scale height remains the same). Figure~\ref{fig:Uphi_Prof_Els_A}b shows that the models with the same $\Lambda(r)$ have near-to identical zonal flow decay in the stable layer, with the zonal winds in models with a stronger imposed dipole strength or a greater electrical conductivity (A1.3 and A2.3 respectively) being quenched most effectively. Although the variation of the background density with radius is rather weak in our models, this suggests that our observations from the Boussinesq models also hold when there is a variable background density. Furthermore, Figure~\ref{fig:Uphi_Prof_Els_A}c, where we plot the ratio of jet amplitude at $0.8r_o$ to that at $r_o$, shows that the $1/\tilde{\rho}(r)$ dependency of the local Elsasser number leads to a more gradual damping of $\overline{u}_\phi$ in these models compared to cases B1. and B2., their Boussinesq equivalents. Figure~\ref{fig:Uphi_Prof_Els_A}c has the same axes as Figure~\ref{fig:Uphi_Prof_Els}c to highlight that these models fit on the same trend line. This is possible for this analysis as the relative decay is evaluated, while the absolute jet amplitudes are difficult to compare with the Boussinesq models.
\section{Discussion and Conclusions}\label{sec:Discussion}

We find that the amplitude and latitudinal extent of zonal flow in the convective region, depends directly on the amplitude of the magnetic forces near the top of the underlying stable region. If these are negligible, due to both a weak dipole field strength and very weak conductivity, the zonal flow at the surface develops a structure and amplitude independent of the magnetic effects acting deep in the stable region below. If Lorentz forces are non-negligible at the bottom of the convective region, they will impact the jets formed above, in particular diminishing those inside the tangent cylinder (see Figures~\ref{fig:UphiDist}d) and \ref{fig:UphiDist_A}c)).\\
The penetration distance of zonal flows into the stable layer is dependent on the product $\sigma B^2$ at depth. For a fixed profile of $\sigma B^2$, it can be expected from \citeA{Wulff_2022} that the degree of stratification, $N/\Omega$, also influences the damping of the zonal winds in the stable layer, as well as other parameters. \citeA{Christensen_2020} suggest that for a fixed $\sigma$ and $B$, and in the limit of negligible viscosity, the combination $(N/\Omega)^2 E_\kappa^{-1}$ is relevant, where $E_\kappa=\kappa/\Omega d^2$ is an Ekman number based on the effective thermal diffusivity in the stable layer. A general scaling relationship for the decay of zonal winds that combines all relevant parameters remains an interesting point for future work.\\
A rough value of the Elsasser number $\Lambda$ at the possible depth in the gas planets where the zonal wind speed is damped may be estimated using interior models and extrapolating the measured axial dipole component downwards. This yields around $\Lambda(0.97R_J)\approx10^{-7}-10^{-5}$ for Jupiter and $\Lambda(0.87R_S)\approx10^{-8}$ for Saturn (with large ranges depending on the models used), substantially less than the values of $\Lambda$ at the points where the winds drop off in our simulations. To compare the wind speeds in our models with those in the two planets we express them in terms of Rossby numbers, i.e. $Ro_\phi=\overline{u}_\phi/\Omega r_o$. On Jupiter $100$~m/s corresponds to a Rossby number of $\sim0.008$, on Saturn $\sim0.01$. For the reference case, the prograde equatorial jet has an amplitude of $Ro_\phi\approx0.01$, the flanking retrograde jets $Ro_\phi=0.004$ and the weakest jets have $Ro_\phi=0.001$, at the surface. Hence the zonal velocities in the simulations are, in terms of their Rossby number, of the right order of magnitude. It is clear that in order to reduce the wind speeds down to the order of cm/s over a short depth as could be inferred from secular variation, the stable layer must be located where the increase in electrical conductivity is sharp. However, as we cannot match the planetary values of parameters related to viscosity or thermal diffusivity in numerical simulations, a quantitative agreement with the plausible reduction of wind speed in the gas planets cannot be expected. \\
When investigating the braking mechanism of the winds in the stable layer, we confirm the findings of \citeA{Christensen_2020} and are also able to explore this further using different models. Firstly, the quenching of $\overline{u}_\phi$ in the stable layer is governed by a thermal wind balance, without magnetic winds playing a role. The temperature perturbation required to facilitate this is generated by meridional circulation in the stable region. Secondly, a significant toroidal field is induced due to the $\Omega$-effect, while the induced poloidal field remains orders of magnitude smaller than the imposed dipole field.\\
Lorentz forces only start acting in the stable region, where small-scale motions are very weak. Therefore they are primarily due to the correlation of the toroidal field, induced by the zonal flows, and the imposed dipole field, as the correlations of the non-axisymmetric field components remain negligible (in contrast to what \citeA{Dietrich_2018} found when varying the radial conductivity profile, without a stable layer). As the Lorentz forces are balanced by Coriolis forces, i.e. the meridional circulation, they indirectly influence the damping of the zonal flows in the stable region. While viscous forces are not dominant in the force balance for the meridional flow, they are not negligible either in our simulations (see Figure~\ref{fig:Meri}), in contrast to what may be assumed in the gas planets.\\
More comprehensive simulations that include the dynamo region \cite{Gastine_2021, Moore_2022} also showed zonal winds inside the tangent cylinder that drop off inside a shallow stably stratified region. Our simpler models, comprising only of the outer regions of the gas planets and imposing a dipolar magnetic field are computationally more economical and allow a more extensive parameter study. Therefore, we are able to compare the influence of varying magnetic parameters and study what factors make the zonal wind damping more efficient. Furthermore, in contrast to \citeA{Moore_2022} our models feature multiple zonal jets, making them more gas planet-like and allowing us to make a systematic study of jet formation and structure.\\
A possible avenue for future work would be to introduce a more complex imposed field at the lower boundary, to study the influence of non-axial-dipole components of the magnetic field, such as intense flux concentrations, similar to Jupiter's observed Great Blue Spot, on the zonal winds.

\section{Open Research}
All simulations were carried out using the 3D magnetohydrodynamic code MagIC which is open source and available at https://magic-sph.github.io/. \add{This specific study requires the feature of an internal axial dipole together with a conducting core which is implemented in version} \citeA{Wulff_2023}. \add{Data to reproduce the figures is available at} \citeA{Wulff_2024}.\\

\appendix
\section{Radial Grid-point Redistribution}

The collocation points are redistributed by the following function:
\begin{equation}
    r=\frac{1}{2}\left[ \alpha_2+\frac{\tan[\lambda(r_{cheb}-x_0)]}{\alpha_1} \right] +\frac{r_i+r_o}{2}
\end{equation}
where $r_{cheb}$ are the Gauss-Lobatto collocation points and $r_{cheb}\in[-1, 1]$. The three parameters are:
\begin{equation}
    \lambda = \frac{\tan^{-1}(\alpha_1(1-\alpha_2))}{1-x_0},\quad x_0=\frac{K-1}{K+1}, \quad K=\frac{\tan^{-1}(\alpha_1(1+\alpha_2))}{\tan^{-1}(\alpha_1(1-\alpha_2))}
\end{equation}
where we use $\alpha_1=2$, $\alpha_2=-0.2$.\\

\acknowledgments

WD would like to thank the Deutsche Forschungsgemeinschaft (DFG), within the Priority Program "SPP 1992 The Diversity of Exoplanets" for their support.
We would like to thank the Isaac Newton Institute for Mathematical Sciences for support and hospitality during the programme DYT2 when work on this paper was undertaken. This work was supported by EPSRC Grant Number EP/R014604/1.

\end{document}